\documentclass[
	a4paper, 
	10pt, 
	unnumberedsections, 
	twoside, 
]{two_col_article}

\usepackage[T1]{fontenc}
\usepackage{comment}
\usepackage[version=3]{mhchem}
\usepackage[utf8]{inputenc}
\usepackage{times}
\usepackage{lipsum}
\usepackage{graphicx}
\usepackage{makecell}
\usepackage{amsfonts}
\usepackage{booktabs} 
\usepackage{amssymb} 
\usepackage{oubraces}
\usepackage[sort&compress,numbers,comma]{natbib}
\setcitestyle{super}
\usepackage{amsmath}
\usepackage{caption}
\usepackage{subcaption}
\usepackage{dblfloatfix}    

\usepackage[ruled, lined, linesnumbered, commentsnumbered, longend]{algorithm2e}

\usepackage[normalem]{ulem}
\newcommand{\markupnote}[1]{{%
        \ifx\showmarkup\undefined%
        \else%
        \color{green}{#1}\color{black}%
        \fi%
}}
\newcommand{\markupdelete}[1]{{%
        \ifx\showmarkup\undefined%
        \else%
        \color{red}\sout{#1}\color{black}%
        \fi%
}}
\newcommand{\markupinsert}[1]{{%
        \ifx\showmarkup\undefined%
        \color{black}{#1}\color{black}%
        \else%
        \color{blue}{#1}\color{black}%
        \fi%
}}
 
\newcommand{\markupaddrow}[1]{%
    \ifx\showmarkup\undefined%
        #1\\*%
    \else%
        #1\\*%
    \fi%
}
\newcommand{\markupdeleterow}[1]{
    \ifx\showmarkup\undefined%
    \else%
        #1\\*%
    \fi%
}
\newcommand{\markupinsertcell}[1]{
        \ifx\showmarkup\undefined%
        #1
        \else%
        \color{blue}{#1}\color{black}%
        \fi%
}
 
\newcommand{\markupdeletecell}[1]{
        \ifx\showmarkup\undefined%
        \else%
        \color{red}\sout{#1}\color{black}%
        \fi%
}
\newcommand{\markupeditcell}[2]{{%
        \ifx\showmarkup\undefined%
        \color{black}{#2}\color{black}%
        \else%
        \color{red}\sout{#1}\color{blue}{#2}\color{black}%
        \fi%
}}
 
\title{
Machine Learning-Assisted Discovery of Flow Reactor Designs
}

\author{%
	Tom Savage\textsuperscript{1,2}, Nausheen Basha\textsuperscript{2}, Jonathan McDonough\textsuperscript{3}, James Krassowski\textsuperscript{3} \\ Omar Matar\textsuperscript{2}, and Ehecatl Antonio del Rio Chanona\textsuperscript{1,2}
}

\date{\footnotesize\textsuperscript{\textbf{1}} Department of Chemical Engineering, Imperial College London\\ \textsuperscript{\textbf{2}} Sargent Centre for Process Systems Engineering, Imperial College London\\ \textsuperscript{\textbf{3}} School of Engineering, Newcastle University}

\renewcommand{\maketitlehookd}{%
\begin{abstract}
Additive manufacturing has enabled the fabrication of advanced reactor geometries, permitting larger, more complex design spaces.
Identifying promising configurations within such spaces presents a significant challenge for current approaches.
Furthermore, existing parameterisations of reactor geometries are low-dimensional with expensive optimisation limiting more complex solutions.
To address this challenge, we establish a machine learning-assisted approach for the design of the next-generation of chemical reactors, combining the application of high-dimensional parameterisations, computational fluid dynamics, and multi-fidelity Bayesian optimisation. 
We associate the development of mixing-enhancing vortical flow structures in novel coiled reactors with performance, and use our approach to identify key characteristics of optimal designs.
By appealing to the principles of flow dynamics, we rationalise the selection of novel design features that lead to experimental plug flow performance improvements of $\sim 60$\% over conventional designs.
Our results demonstrate that coupling advanced manufacturing techniques with `augmented-intelligence' approaches can lead to superior design performance and, consequently, emissions-reduction and sustainability.
	\end{abstract}
}

\begin{document}
\maketitle
\thispagestyle{empty}
\section{Introduction}

\begin{figure*}[htb!]
    \centering
    \includegraphics[width=\textwidth]{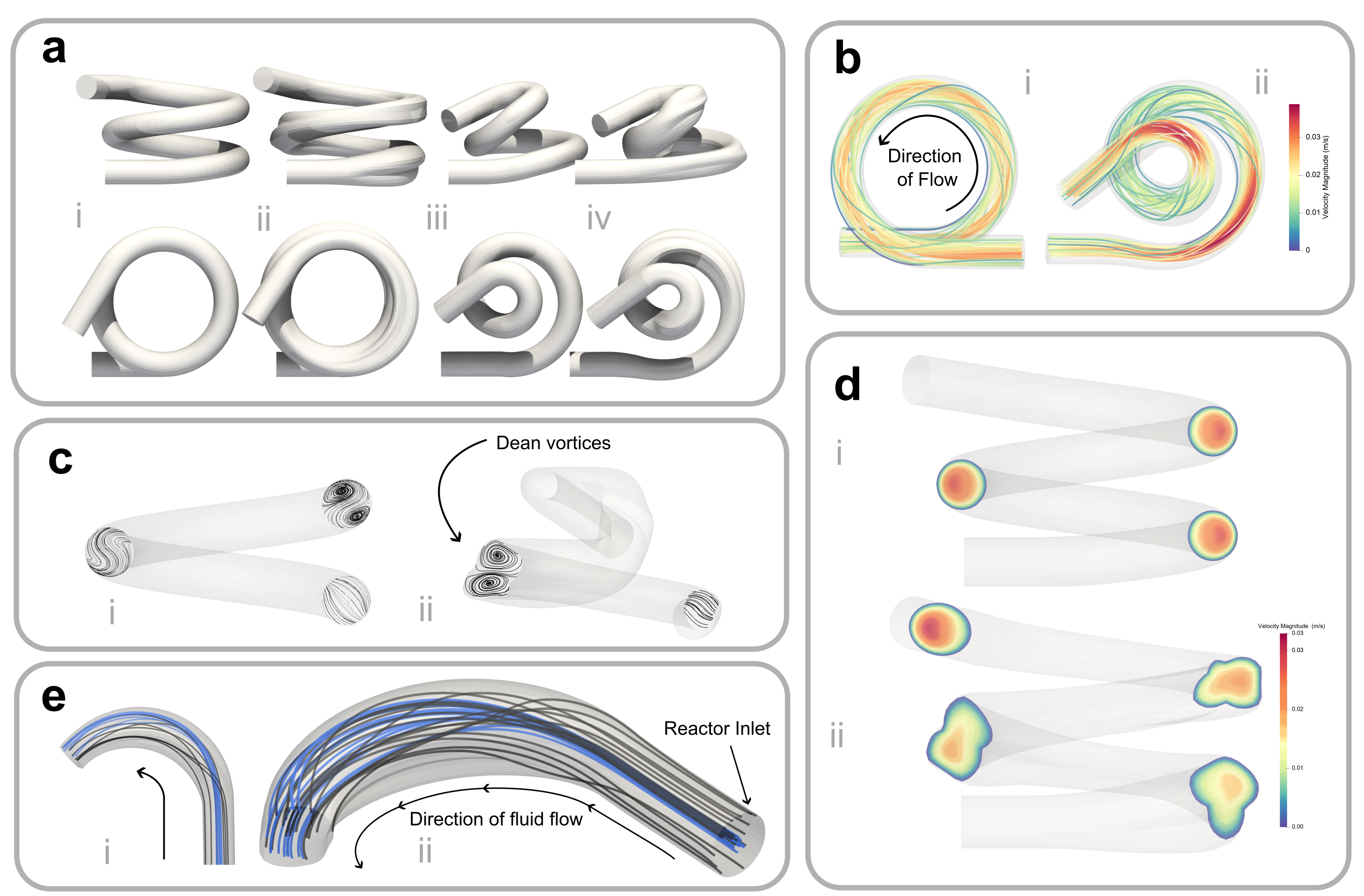}
    \caption{\textbf{Overview of optimal reactor characteristics.} \textbf{a}, The conventional coiled-tube reactor (i) alongside optimal coiled tube reactors generated by parameterising the cross-section (ii), the coil path (iii), and a joint parameterisation (iv). \textbf{b}, Velocity streamlines coloured with velocity magnitude within a standard coil (i) and within optimal joint parameterisation (ii). \textbf{c}, Secondary flow streamlines at various cross-sections of the coil. \textbf{d}, Cross-sectional plane across the coil demonstrating streamwise velocities \textbf{e}, The presence of induced Dean vortices within the optimal joint parameterisation coil (ii) compared with a standard coil (i).}
    \label{fig:overall_fig}
\end{figure*}

Advances in additive manufacturing have enabled the fabrication of a wide range of complex and potentially counter-intuitive reactor designs. 
Previously infeasible, or highly impractical designs can now be manufactured and investigated, resulting in substantially larger design spaces. 
Coupled with data-driven design tools, such as multi-fidelity Bayesian approaches\citep{savage2023multifidelity,BOAH,He2017,folch2023combining,takeno,Lam2015,mcleod}, which have enabled the optimisation of large-scale simulation-based problems, a pathway has emerged towards the identification of optimal reactors from larger design spaces with the potential for improved performance. 
By exploiting lower-fidelity simulations throughout optimisation, high-quality solutions can be generated in a significantly reduced time; this is particularly the case in scenarios where gradients are unavailable or a more global solution is desired than gradient-based approaches\citep{savage2023multifidelity,takeno}.

The purpose of this article is to demonstrate a general `augmented-intelligence' framework that enables the design of new superior reactors that surpass the performance of conventional designs.
To do so, we take advantage of developments in data-driven optimisation, machine learning, computational fluid dynamics, and additive manufacturing; the coiled tube reactor was chosen as an illustrative exemplar.

Coiled tube reactors have received attention across chemical engineering due to their desirable mixing and heat transfer characteristics \citep{McDonough2019coilincoil,McDonough2019a,Agrawal2001,Jokiel2019,Wang2021, Pukkella2022}. 
Their applications range from flow chemistry\citep{Porta2015} and bioprocesses\citep{Hagedorn1990}, to chemical kinetic experiments \citep{Jokiel2019}. 
At the mesoscale, coiled tube reactors have been shown to combine the heat and mass transfer properties of microreactors with the economic benefits of high-throughput larger scale reactors\citep{Dong2022,grande2020multiscale}.
Previous work has revealed improvement in the plug flow performance of coiled tube reactors at relatively low flow rates (characterised by Reynolds numbers $Re\leq50$) by super-imposing pulsed-flow operating conditions, which induce Dean vortices\citep{basha,McDonough2019a,Nivedita2017} that enhance radial mixing; in the absence of pulsed flow, such vortices develop at much higher flow rates and Dean numbers ($Re > 300$, $De >75$)\citep{Dean1927,Gao2023, ligrani_1994_a}. 
It is essential to induce the formation of Dean vortices at low flow rates and Dean numbers, over large proportions of the reactor interior, under steady-flow conditions, and without the added overhead of pulsed-flow forcing.

We seek to enhance the plug flow performance of coiled-tube reactors by identifying two novel parameterisations for the reactor geometry in the radial and axial directions.
We solve the simulation-based optimisation problem for all parameterisations using multi-fidelity Bayesian optimisation, considering steady-flow conditions at a low flow rate (with $Re=50$ for which pulsed-flow would have been necessary to drive vortex formation). 
The composite objective we maximise consists of plug flow performance, which we approximate from computational residence-time distributions using a tanks-in-series model, and a non-ideality term that penalises bimodal or unsymmetrical distributions.
Optimal solutions are investigated, where, to the best of our knowledge, we identify the presence of fully-developed Dean vortices for the first time at low Reynolds numbers under steady-flow conditions.
The key driving factors that result in improved performance are identified from which we design and present two reactors.
We 3D-print and experimentally validate these designs, confirming their improved performance over a conventional coiled tube reactor across both tracer and reacting flow experiments.

Our work establishes a framework for the design of next-generation reactors that can significantly improve the performance, sustainability, and economic viability of various manufacturing processes. 
Ultimately, this work aims to shift the paradigm of reactor design to take advantage of the suite of modern computational methodologies in design and optimisation, demonstrating new opportunities to support the discovery, innovation, and advancement across chemical engineering.

\section{Results}\label{results}
Each parameterisation was optimised under steady flow conditions, with a Reynolds number of 50. 
Details of both parameterisations can be located within the Methods.
To ensure tractability of the problem, the joint parameterisation employs a sequential strategy whereby the parameters of the optimal coil path design were fixed before the introduction of parameters to manipulate the cross-section. 
\textbf{Figure 1a} demonstrates the most optimal geometries for each parameterisation.
We denote the control reactor as design `i', and designs resulting from optimal cross-section, coil path, and combined parameters as `ii', `iii', and `iv', respectively, corresponding to the subplots within \textbf{Fig. 1a}.

\subsection{Effect of Design Features on Flow Structures}
We first consider design `ii', where the shape of the tube cross-section throughout the length of the reactor is allowed to vary.
The first key feature of design `ii' is that the cross-section undergoes periodic expansions and contractions approximately every half-turn.  
Secondly, the design comprises a pinch constricting the flow when the cross-sectional area is greatest, during the expansion phase. 
Next, we consider the optimal solution resulting from the parameterisation of the coil path corresponding to design `iii' where the path deviates from a nominal configuration via the interpolation of cylindrical coordinates.
The coil radius of curvature of design `iii' begins relatively large, and subsequently reduces along the length of the reactor, resulting in a tighter design than that in `ii'.  
The pitch of the coil in design `iii' begins small before rising approximately halfway along the length of the reactor then decreasing near the reactor outlet to the extent that the coil path points  downwards.  
Within design `iv', the path is fixed as in `iii' and the cross-section is allowed to vary; inspection of design `iv' reveals that it possesses features that are present in both designs `ii' and `iii'.
To further investigate why these solutions are deemed optimal, we demonstrate different flow characteristics. 
\textbf{Figure 1b} depicts changes in fluid velocity within design `iv' compared to a conventional coil (design `i').
The expansion and contraction features in the cross-section alter the velocity distribution along the coil's length in design `iv', resulting in higher and lower velocities during contraction and expansion, respectively.
Conversely, the conventional coil exhibits a relatively uniform velocity distribution along the coil length. 
The velocity changes observed in design `iv', due to acceleration and deceleration, induce stronger pressure gradients, leading to the formation of Dean vortices that significantly enhance radial mixing.
\textbf{Figure 1c} demonstrates the formation of Dean vortices in both a standard coil and design iv, as shown by sub-plots i and ii in  \textbf{Figure 1c}, respectively. 
While Dean vortices are also formed in a standard coil, they are only partially established close to the tube outlet. 
The earlier formation of fully-developed Dean vortices in design `iv' demonstrates the impact of reactor geometry resulting from the application of our framework, suggesting the potential for enhanced plug flow performance within more compact reactors, at lower $Re$.

Promoting plug flow depends on the relationship between axial and radial fluid mixing. 
To control axial dispersion, the consideration of streamwise velocity distribution becomes imperative. 
Therefore, we present velocities across the cross-sectional area of the coil.
\textbf{Figure 1d} represents the velocity distribution along the length of design ‘i' and design ‘iv', illustrated by sub-plots i and ii, respectively.
In design ‘i,’ we observe that the peak velocity is closer to the outer walls, with lower velocities near the inner wall throughout the coil length. 
This configuration induces faster tracer movement along the outer walls and slower movement along the inner walls, resulting in an extended time for the tracer to exit the coil, and increased axial dispersion.  
In contrast, the inclusion of the characteristic pinch feature in design `iv' plays a key role in redistributing velocity across the coil cross-section by altering the radial position of peak velocities along the coil length. Design 'iv' demonstrates peak velocity at various radial positions along the coil's length: at the top, bottom, and towards the outer wall, as depicted in \textbf{Figure 1d}. 
This leads to the acceleration and deceleration of the tracer movement along the coil length, promoting a comparatively uniform axial movement of the tracer over time. 
Consequently, this results in a narrow profile for the residence time distribution of the tracer.

To further illustrate the superior characteristics of the optimised designs over conventional coil tube reactors, \textbf{Figure 1e} depicts streamlines representing fluid flow within the initial length of design `ii'.
Blue streamlines originate from the coil centre, while black streamlines start near the coil wall.
As the cross-section of the reactor expands throughout the initial curve, the slow-moving fluid radially furthest from the centre of the coil moves initially outwards. 
Subsequently, as the cross-sectional area is greatest, the fluid within the central region moves towards the outer walls of the coiled tube. 
The fluid closest to the outer walls is then acted upon by the change in cross-section of the tube in the form of a pinch forcing fluid towards the centre of the tube via a swirling motion. 
The cross-section then contracts again, enabling a repetition of the mechanism for radial mixing under steady-flow conditions.

\subsection{Convergence Analysis}

\begin{figure*}[htb!]
    \centering
        \includegraphics[width=\textwidth]{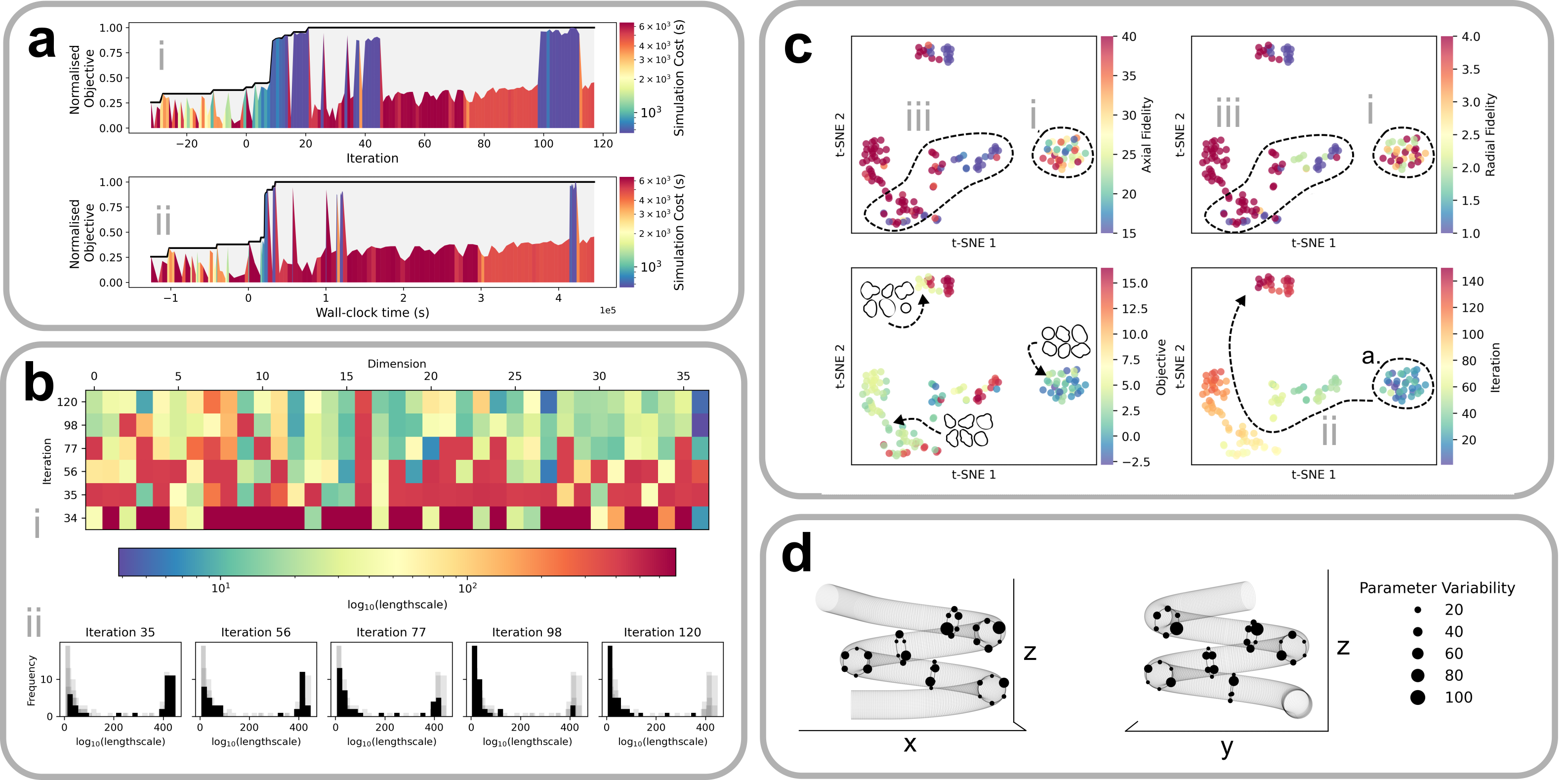}
         \caption{\textbf{Analysis of residence time distributions and optimisation convergence.} \textbf{a}, The optimisation objective against iteration (i) and wall-clock time (ii). The objective has been normalised. The initial data set generated via design-of-experiments is denoted as negative iterations and wall-clock time. Hence, optimisation begins at the 0\textsuperscript{th} iteration and at $t=0$. \textbf{b}, Gaussian process dimension lengthscales throughout Bayesian optimisation iterations (i) alongside histograms demonstrating the distribution of GP lengthscales changing throughout optimisation (ii). \textbf{c}, t-SNE analysis of the data generated throughout optimisation in design parameter space ($\mathcal{X}$), reducing the dimensionality of design parameters to two dimensions, and labelled with different respective quantities. \textbf{d}, Parameter variability, defined as a function of the GP lengthscale corresponding to each parameter, plotted for each inducing parameter on a nominal coil.}
\end{figure*}

In this section, we focus on the convergence of the designs from an optimisation perspective in order to highlight the potential optimality of the solutions provided. 
Gaussian processes (GPs) are used to model the simulation cost and objective throughout the design space, and are iteratively updated based on simulations selected through a multi-fidelity acquisition function.
Full details of the optimisation can be found within the Methods.
Given the high-dimensional design space, we apply t-Distributed Stochastic Neighbour Embedding (t-SNE) to identify emerging trends throughout optimisation, investigate design convergence, and highlight the behaviour of the GP hyperparameters (see Methods). 
This is a probabilistic dimensionality reduction technique, enabling high-dimensional data to be plotted whilst preserving local structure. 
We focus on the convergence of the cross-section design problem, with the trends associated with the other parameterisations relegated to the Supplementary Information.
We find that the cross-section optimisation problem has better convergence properties than the other two design problems, motivating our methodology in identifying optimal driving characteristics as opposed to designs themselves.
\textbf{Figure 2a} demonstrates the composite objective function against iteration (i), and wall-clock time (ii), demonstrating how lower-cost simulations are applied to provide less-expensive exploration. 

The initial design of experiments used within optimisation is represented as occurring before the first iteration (so-called `negative' iterations), and before $t=0$.
This sample contains reactor simulations with a variety of computational costs, as simulations across the spectrum of fidelities are performed. 
After optimisation begins, low-cost, low-fidelity simulations are selected during an initial phase of exploration. 
Simulations are selected that are progressively less expensive until approximately the 20\textsuperscript{th} iteration, as the updated GPs within the framework gain a better representation of simulation cost, and are able to select more efficient simulations.
The framework then undergoes a period largely applying either the highest fidelity simulations, or the lowest fidelity simulations. 

To investigate the properties of the GP that model the objective throughout optimisation, we observe the kernel function hyper-parameters at a number of iterations. 
Figure \textbf{2bi} demonstrates the lengthscale of each dimension within the kernel function within the objective GP as optimisation progresses. 
Initially, the majority of lengthscales are relatively large indicating a uniform function space, capturing broad trends, as expected in a low-data regime in high dimensions. 
As optimisation progresses, and the GP is trained with more data, the lengthscales broadly decrease indicating an improved representation of the design space.
Figure \textbf{2bii} demonstrates the distribution of lengthscale hyper-parameters as optimisation progresses. 
The distribution of hyper-parameters tends towards lower values, and the GP becomes more nonlinear as data become available and correlations are more accurately captured.

As different simulation fidelities bias the objective function and the parameterisations are high-dimensional, we apply t-SNE to the convergence data of the cross-section parameterisation, observing data in the space of design parameters but not fidelities, colouring data points based on number of properties. 
Figure \textbf{2c} shows the two-dimensional t-SNE embeddings of data points $\in \mathcal{X}$, labelled by axial and radial fidelities, objective value, and iteration. 
Feature \textbf{i} shows the initial data set generated before optimisation begins. 
From here, the optimisation can be seen to progress along the path denoted by \textbf{ii}, demonstrating systematic changes in parameters indicative of convergent behaviour.
Along this path, a number of different fidelities are evaluated (feature \textbf{iii}), demonstrating how lower-fidelity simulations are applied. 

We present a sample of the inducing cross-sections from various clusters of data within the analysis against optimisation objective. 
The cross-sections in the original data set (feature \textbf{i}), as expected contain no distinct characteristics. 
As the optimisation progresses, the cross-sections defined by inducing points become more distinct, with certain cross-sections gaining symmetrical forms.
Finally, at the end of optimisation, the cross-sections are most distinct, consisting of alternating small and large cross-sectional areas, with pinches throughout. 

Lastly, we define parameter variability as the normalised inverse of each GP lengthscale, each corresponding to a specific parameter.
We calculate this property at the end of optimisation and present the value for each inducing parameter (the collection of which defines the form of the design space), indicating the variability in objective each specific parameter is responsible for. 
As can be seen from Figure \textbf{2d}, which demonstrates parameter variability throughout the reactor, the variability resulting from parameters in the tube cross-section is greater where pinches occur, than those associated with the bottom and top of the tubular cross-section. 
These trends indicate that the tube cross-section, and specifically the pinch, has a significant effect on plug flow performance, confirming our observations.

\subsection{Optimal Designs \& Comparison with 3D-Printed Reactors}

\begin{figure}[htb!]
    \centering
    \includegraphics[width=\columnwidth]{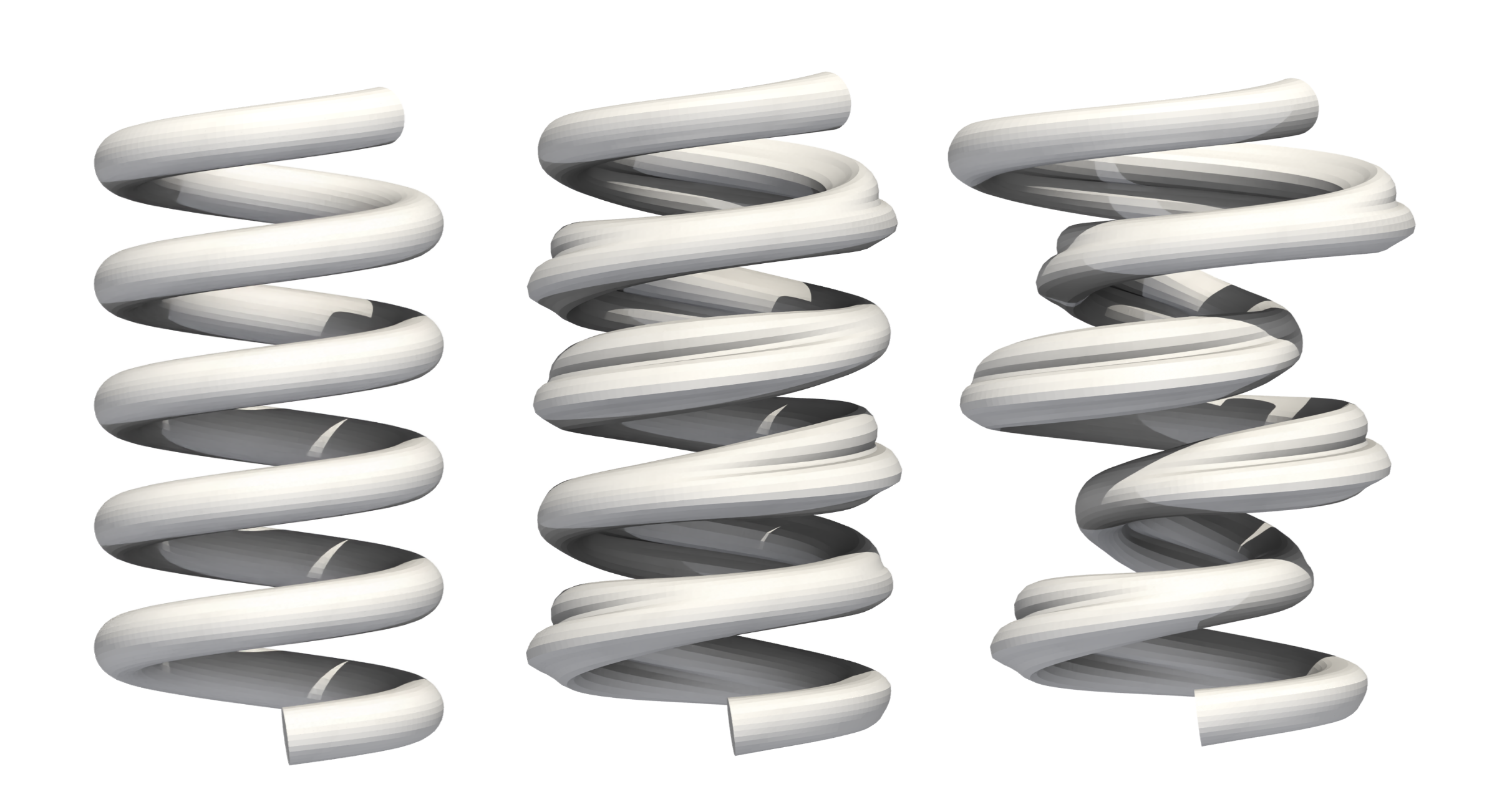}
    \caption{The nominal coiled tube (\textbf{left}, R1) alongside extrapolated steady-flow coil designs containing aspects from the optimal cross-section (\textbf{centre}, R2) and both cross-section and coil path (\textbf{right}, R3). }
    \label{fig:experimental}
\end{figure}

\begin{figure*}[htb!]
    \centering
    \includegraphics[width=\textwidth]{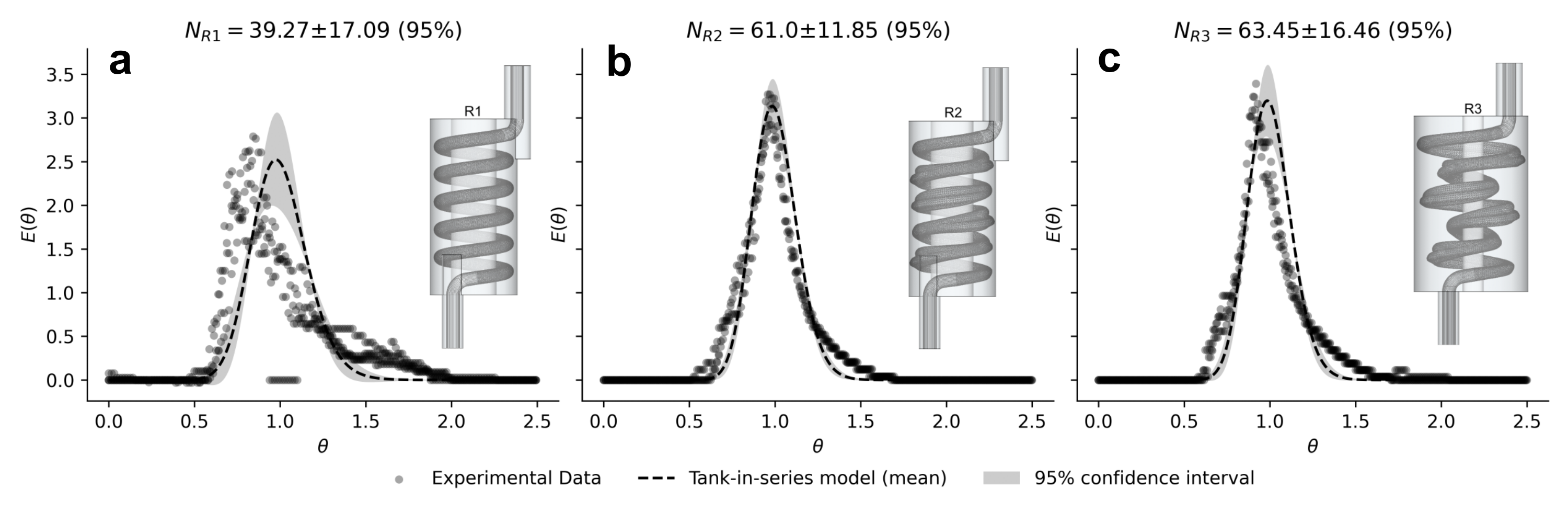} \caption{\textbf{Data generated across three experiments for each reactor configuration, designed to maximise equivalent tanks-in-series and minimise non-symmetric RTDs. A tanks-in-series model is used to estimate the performance of each reactor across the experiments performed. Dimensionless concentration at the outlet of the reactor ($E(\theta)$) is plotted against dimensionless time ($\theta$).} \textbf{a}, the control reactor (R1). \textbf{b}, the reactor with variable cross-section (R2). \textbf{c}, the reactor with variable cross-section and path (R3).}
    \label{fig:experimental_3d}
\end{figure*}

In the foregoing, we have shown that the design concepts identified through the use of data-driven optimisation and machine learning are as follows: 
\begin{itemize}
    \item Expansion and subsequent contraction of reactor cross-section every half turn;
    \item Distinct pinch in cross-section throughout the reactor;
    \item Changes in the direction of flow including constriction of coil radius.
\end{itemize}
Previous work has also demonstrated the existence of a linear relationship between the number of coils and plug flow performance\citep{McDonough2019b}. 
We now apply the design concepts to two reactors corresponding to longer coils than those studied above; for one, we include changes to the tube cross-section, and for the other,  variations to both cross-section and coil path. 
\textbf{Figure \ref{fig:experimental}} demonstrates the final coiled-tube reactor designs, containing features identified as being responsible for the optimality of both the cross-section and coil path parameterisations.

The designs shown in Figure \ref{fig:experimental} were 3D-printed, and residence-time distribution experiments (see Methods) were performed for each configuration using steady-flow at a $Re=50$. 
\textbf{Figure \ref{fig:experimental_3d}} demonstrates the aggregated experimental data alongside tank-in-series models for each reactor configuration.
The standard coil (R1) produces a skewed distribution with a mean equivalent tanks-in-series value across all experiments of of 39.27. 
The two distributions resulting from proposed reactors with variable cross-section (R2) and cross-section with coil path (R3) are symmetrical.
The equivalent tanks-in-series for these two reactors are higher than the standard coil at 61 (R2) and 63.45 (R3), representing a 55\% and 62\% increased value of mean equivalent tanks-in-series respectively.
The RTD resulting from R1 is to be expected, as typically high $Re$ is required for Dean vortices to form. 
The improvements in R2 and R3 demonstrate that we can avoid the need for oscillatory conditions to induce Dean vortices and promote mixing at low $Re$, through the inclusion of the identified design concepts.
Other factors such as multiphase flows and chemistry may be explored to further exploit the discovery of designs through machine learning.
An uncertainty analysis of the experimental data can be found in the Supplementary Information.

To further validate the effectiveness of the optimized reactor designs, we performed experiments using the Villermaux–Dushman reaction system \citep{Pinot2014}, a widely used test reaction for characterizing mixing performance in reactors. The Villermaux–Dushman reaction is particularly suitable for this purpose as it is a fast, mixing-sensitive reaction that can provide insights into the mixing efficiency of the reactors\citep{McDonough2019mm}.
The Villermaux-Dushman reaction system is as follows:

\begin{align}
    \cee{H2BO^-3 + H^+ &<=> H3BO3}\\
    \label{reaction2}\cee{IO3^{-} + 5I- + 6H+ &<=> 3I2 + 3H2O}\\ 
    \cee{I2 + I- &<=> I3^-}
\end{align}

Under poor mixing conditions reaction \ref{reaction2} is prohibited by the lack of acid, consumed within the faster first reaction. 
Good mixing conditions results in the formation of \ce{I2} which in turn reacts to produce \ce{I3-} tri-iodide ions, measured at 353nm \citep{McDonough2019mm}. 

\begin{table}[htb!]
\caption{Absorbance values at 353 nm at the outlet of each reactor following a Villermaux–Dushman reaction.}\label{reaction_table}
\begin{tabular}{lllll}
\hline
 & \multicolumn{4}{c}{\textbf{Absorbance (353 nm)}} \\ \hline
Reactor & Repeat 1 & Repeat 2 & Repeat 3 & Average \\ \hline
R1 & 0.267 & 0.276 & 0.275 & 0.2727 \\
R2 & 0.292 & 0.288 & 0.300 & 0.2933 \\
R3 & 0.326 & 0.324 & 0.325 & \textbf{0.325} \\ \hline
\end{tabular}
\end{table}

\textbf{Table \ref{reaction_table}} presents the absorbance values at 353 nm, corresponding to the presence of tri-iodate, measured at the outlet of each reactor. 
A higher absorbance value indicates a higher conversion of the limiting reagent.
The optimized reactors, R2 and R3, exhibit higher average absorbance values (0.2933 and 0.325, respectively) compared to the nominal design, R1 (0.2727), demonstrating improved reactive flow performance at Re=50.
These results confirm that for mass transfer-limited reactions, such as the Villermaux–Dushman reaction, optimizing the residence time distribution serves as an effective proxy for enhancing overall reactor performance.

We note that the power consumption of the experimental syringe pumps was not measured.
However, the pressure drop between the standard and optimised coils can be analysed computationally via CFD simulations. 
For a single coil turn, the optimised geometry has a 29.3 \% higher pressure drop compared to the standard coil. 
This difference would further increase with the coil length. 
Therefore, the proposed designs almost certainly have a larger pressure drop than the standard coil.
Future work may propose to treat this, or a similar case as a multi-objective black-box optimisation problem, trading off pressure-drop and plug flow performance.

\section{Discussion}

Throughout optimisation, we cannot guarantee that the global optimum of the parameterisations has been identified.
However, the global methodology we apply provides more varied solutions than a gradient-based local approach such as the adjoint method, and we identify the key features that drive the behaviour of fluid within these reactors, including the presence of induced Dean vortices. 
Subsequently, the reactors that we experimentally validate can be interpreted as `augmented intelligence'-assisted designs, with features based on the optimal solutions. 
We propose this workflow for the design of highly-parameterised reactors as the interpretability of solutions is maintained. 
To support the design of algorithms for the optimisation of high-dimensional expensive black-box problems, we release all parameterisations as benchmark problems\footnote{Found at \url{https://github.com/trsav/reactor_benchmark}}.
The parameterisations we optimise contain feasible reactor geometries by design, resulting in an unconstrained optimisation problem. 
Highlighting the importance of parameterisation specification, the emergent behaviour in designs `iii' and `iv' contains a coil path that unintendedly pitches down near the reactor outlet. 
It is worth noting that features of the optimal coil path design bear some resemblance to the wavering coiled flow inverter  design proposed by \citet{Singh2019}, though this design primarily relies on inversions within the coil.
Future work may perform comparisons over a wide range of alternate and existing coil designs.

We anticipate that the workflow for machine-learning-assisted discovery we present here can be used across a large number of expensive simulation-based design and optimisation problems involving, for instance, reactive and multi-phase flows. 
Alongside developments in additive manufacturing, we hope that chemical reactors discovered via machine learning become increasingly prevalent to support future sustainable processes.

\section*{Methods}\label{methodology}

\subsection{Parameterisations}
In this section, we focus on the design of efficient parameterisations for the discovery of novel coiled-tube reactors, ensuring tractability of the design problem. 
We present two distinct parameterisations: one where the geometry of the cross-section varies along the length of the reactor, and another where the reactor path itself is allowed to vary. 
To manage the flexibility/viability trade-off, both parameterisations themselves are defined by hyper-parameters determining their complexity.
To ensure smooth transitions in these parameterisations, we employ interpolating points in their formulation, both to define the reactor path, and the cross-section throughout the reactor length.
We first introduce Gaussian processes in polar coordinates as a means to generate variable tube cross-sections from interpolating points.

\subsubsection{Polar Gaussian Processes}

A Gaussian processes is an infinite-dimension generalisation of a multi-variate Gaussian distribution \cite{williams2006gaussian}. 
The mean vector and covariance matrix are replaced by a mean function and kernel function, respectively. 
A Gaussian process can be described as
\begin{align*}
    f(x) \sim \mathcal{G}\mathcal{P}(m(\mathbf{x}),k(\mathbf{x},\mathbf{x}^{'})).
\end{align*}
The kernel function $k$ dictates the behaviour of functions from this distribution, and can be parameterised by hyper-parameters including length scale.
By conditioning a Gaussian process on a data set $\mathbf{X}_*$, a posterior distribution of functions can be obtained. 
At inputs $\mathbf{X}$, and previously evaluated function values $\mathbf{y}$ the posterior predictive mean and standard deviation become
\begin{align*}
    \mu_f(\mathbf{X})&=K(\mathbf{X},\mathbf{X}_*)K(\mathbf{X},\mathbf{X})^{-1}\mathbf{y} \\ 
    \sigma_f(\mathbf{X}) &= K(\mathbf{X}_*,\mathbf{X}_*)-K(\mathbf{X}_*,\mathbf{X})K(\mathbf{X},\mathbf{X})^{-1}K(\mathbf{X},\mathbf{X}_*)
\end{align*}
where $K$ is a covariance matrix derived from kernel function $k$.

\begin{figure*}[htb!]
    \centering
    \includegraphics[width=\textwidth]{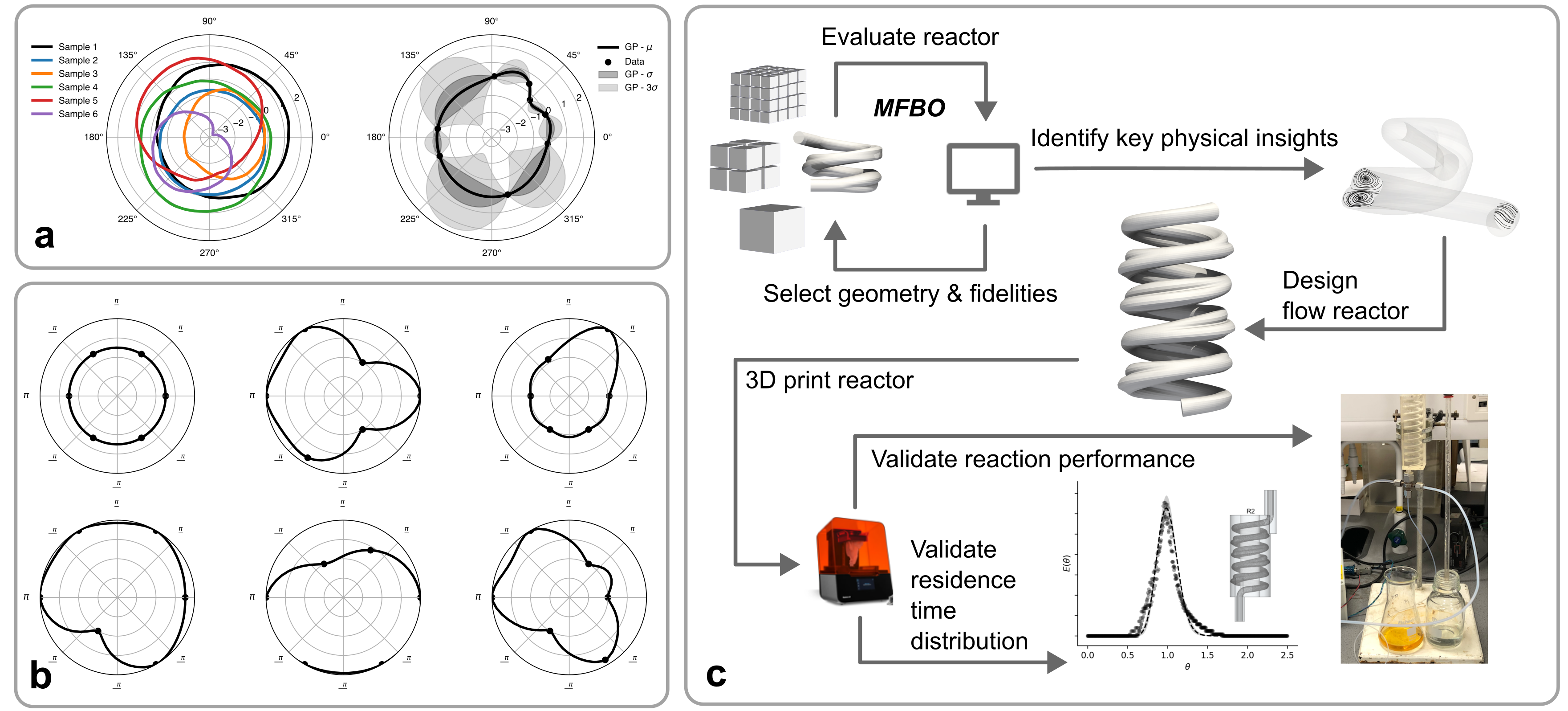}
    \caption{\textbf{Aspects of the methodology.} \textbf{a}, \textbf{Left}: Polar Gaussian process parameterisation of coil cross section, demonstrating samples from a Gaussian process prior with a polar kernel. \textbf{a}, \textbf{Right}: The posterior distribution after the polar Prior distribution has been conditioned on data. In this demonstration we assume noiseless observations. \textbf{b}, Examples of sets of inducing points at given locations along the length of a coil. A polar Gaussian process is used to interpolate between points, where each data point's radial value is a parameter. \textbf{c}, A flow-diagram of the main aspects of the methodology of this article.}
    \label{polar_gp}
\end{figure*}

The squared-exponential kernel function assigns decreasing correlation between locations in input space with increasing Euclidian distance, providing an intuitive interpretation for the majority of regression settings. 
Other kernel functions have been proposed to provide valid covariance matrices, resulting in Gaussian process prior distributions with different properties. 
The polar squared exponential kernel enables valid covariance matrices to be constructed in polar coordinates and was outlined by \cite{padonou:hal-01119942}.
A standard kernel function, dealing with data within data in polar coordinates, will determine that data at $\theta_1=0$ and $\theta_2=\pi/2$ are highly uncorrelated. 
This is untrue, and in the presence of noiseless-observations these two data points should have perfect correlation ($k(\theta_1,\theta_2) = 1$).
Polar kernel functions enable smooth interpolation in polar coordinates by including the ability for proper distances and respective covariances to be calculated between any two angles \cite{Pinder2022,padonou:hal-01119942}. The polar covariance function is written as 
\begin{align}
   k(\theta,\theta^{'}) = \left|\left(1+\tau \frac{d(\theta,\theta^{'})}{\pi}\right)\left(1- \frac{d(\theta,\theta^{'})}{\pi}\right)^\tau\right| \quad \tau \geq 4.
\end{align}
The angular distance metric $d$ is given as
\begin{align*}
d(\theta,\theta^{'}) = |(\theta-\theta^{'}+\pi) \mod 2\pi - \pi |,
\end{align*}
where $\tau$ is a hyper-parameter analogous to length-scale and controls how smooth the prior distribution of functions is. 
Figure \ref{polar_gp}a demonstrates samples from a Gaussian process prior with polar kernel, as well as a Gaussian process with polar kernel posterior distribution conditioned on data. 

\subsubsection{Coil cross-section Parameterisation}

Initially, we define the number of interpolating points for a given tubular cross-section, denoted $n_c$, and indexed by $i$.
We then distribute these points equally along the angle $\theta$ in polar coordinates.
The radius coordinate, $r_i \in n_c$ for each inducing point serves as the decision variable, or parameter.
Given a set of inducing points corresponding to a specific cross-section, a polar Gaussian process is used to interpolate between them, resulting in a valid, continuous curve in polar coordinates defining an individual cross-section. 
Next, we establish the number of interpolating cross-sections denoted $n_l$ and indexed by $j$, equally spaced along the length of the coil.
Both $n_c$ and $n_l$ represent hyper-parameters; increasing either one raises the dimensionality of the resulting design problem, whilst resulting in a more flexible parameterisation. 
Figure \ref{polar_gp}b demonstrates the polar GP interpolation of each specific cross-section along the length of the coil, with $n_c = 6$ and $n_l = 6$.
To ensure compatibility with existing fixtures and fittings, two additional cross-sections are defined at the beginning and end of the coil, with constant radius.

Subsequently, we place each cross-section along the defined reactor path in 3D space and rotate each cross-section to face perpendicular to the direction of flow. 
The resulting interpolated cross-sections are defined by a quadratic interpolation between each polar GP posterior, in cylindrical coordinates.
Finally, an inlet and outlet is added to the reactor by extending the cross-sections at the beginning and end of the reactor. 

The data generated from the parameterisation is provided to code to mesh tubular reactors\footnote{Found at \url{https://github.com/OptiMaL-PSE-Lab/pulsed-reactor-optimisation}}.
Importantly, the meshing procedure enables control over the number of cells throughout the axial direction of the reactor (the direction of fluid flow), as well as the radial direction. 
Defined by axial and radial fidelity parameters respectively, both values dictate the factor to which blocks are subdivided during mesh creation.
By maintaining the ability to adapt the fidelity of the mesh in two independent directions, we provide greater scope for identifying efficient (regarding information gained on a computational expense basis) simulations. 
Previous work has demonstrated how the output of coiled-tube reactor simulations varies approximately smoothly with changes in fidelities \cite{savage2022,savage2023multifidelity}.
In this article we assume them to be continuous throughout modelling and optimisation, and round them to the nearest integer when stored or evaluated.
We direct the reader to previous work \citep{savage2023multifidelity}, where various fidelities are validated against experimental data confirming the validity of the high-fidelity model with respect to cell-count and the relationship between residence-time distributions for lower fidelity simulations is explored.

\subsubsection{Coil Path Parameterisation}

To achieve a parameterisation of the coil path, we start by defining a baseline path in cylindrical coordinates, $(\rho_0, \phi_0, z_0)$. 
This path is based on a standard coil configuration and serves as a reference for introducing induced deviations. 
Let $\Delta \rho_j$ and $\Delta z_j$ be the deviations in the radial and height directions, respectively, for each inducing point $j$. 
In optimising for inducing point deviations, we ensure that we maintain the coil's non-intersection constraint by design.

We introduce the decision variables as mismatch terms to the baseline path, resulting in a parameterised path given by $(\rho_j, \phi_j, z_j) = (\rho_0 + \Delta \rho_j, \phi_0, z_0 + \Delta z_j)$. 
After defining the path, we place circles defining the constant cross-section in 3D space parallel to the direction of flow. 
Let $n_p$ be the number of inducing points along the coil path, indexed by $j$. 
This parameterisation is similar to the one presented by \cite{Cimolai2022}.
We define 6 inducing points for both the axial direction and radial direction, i.e. $n_p=6$, however none in the rotational axis $\theta$. 
Deviations in $\theta$ may be included, however we find that this often results in self intersections when the coil cross-section is defined from the path. 

Further details regarding the implementation of all parameterisations can be found within the supplementary information. 

\subsubsection{Nominal Reactor and Parameter Bounds}

We first specify a nominal reactor, the path of which is used across both parameterisations. 
Based on previous work, the standard coiled-tube reactor performs optimally with a low pitch, and large coil radius \citep{savage2023multifidelity}. 
We specify the nominal pitch, $p_n$ as 10.4mm and nominal coil radius, $C_{r,n}$ as 12.5mm. 
Previous work has demonstrated that as the number of coils increase, the number of equivalent tanks-in-series rises approximately linearly \citep{McDonough2019a}. 
Therefore we specify two coil rotations, balancing overall simulation time and reactor size, providing an overall reactor length of $4\pi C_{r,n}$.
For cross-sectional interpolating points we apply lower and upper bounds of 2mm and 4mm respectively, indicating minimum and maximum radial values.
To maintain feasible coil paths we define the bounds of height deviations $\Delta z$ as $\pm 1$mm, and radial deviations $\Delta \rho$ as $\pm 3.5$mm, ensuring no self-intersections.

\subsection{Simulation}

We apply an established methodology for the evaluation of reactor designs under specific operating conditions, similar to the approach used in our previous work \citep{savage2023multifidelity}. 
Utilizing OpenFOAM, a simulation is performed where an impulse tracer is injected into the water medium, and the resulting concentration is tracked by solving the relevant transport equations. 
Key aspects of the approach include the application of the transient pimpleFOAM solver for unsteady momentum equations, and the scalarTransportFoam to handle convection-diffusion effects with diffusion coefficient constant at a low value.
The integral of the concentration over the outlet is recorded at each timestep and returned from the simulation.
The evaluation process additionally involves monitoring the tracer concentration at the outlet to terminate the solution at a specific threshold, ensuring minimal solution times.
This method has been adapted to the current study, enabling the evaluation of highly-parameterised reactors, more information can be located within \citep{savage2023multifidelity} and \citep{basha}.

\subsection{Optimisation}

The optimisation of highly-parameterised reactor simulations is high dimensional and involves computationally expensive function evaluations. 
To solve this problem we apply multi-fidelity Bayesian optimisation.
The specific approach we apply, DARTS, was presented by \cite{savage2023multifidelity} for the optimisation of simulated reactors and tubes. 
The method enables multiple continuous simulation fidelities to be considered simultaneously. 
In addition the approach explicitly models the cost of a simulation as a function of simulation fidelity as well as inputs (such as geometry parameters).

\subsubsection{Multi-fidelity Bayesian optimisation}

Expensive derivative-free optimisation problems, such as the optimisation of CFD simulations, selection of appropriate neural network hyper-parameters, or the optimal design of experiments exist throughout a number of domains.
Bayesian optimisation has been proposed to address these challenges, with the aim of determining optimal solutions within an tolerable computational or time-budget. 
The majority of computationally expensive functions share the common trait: their complexity can be reduced at the expense of accuracy, resulting in the ability to perform function evaluations with the same set of decision variables with varying degrees of confidence, dictated by one or more fidelities.
Fidelities control both the bias and noise of function evaluations.
For example the number of discrete cells used for the evaluation of a flow-field, reducing which results in a less accurate output.
Leveraging lower-fidelity function evaluations for the optimisation of expensive systems in which there is a time or computational budget required to identify an optimal solution can improve or enable the solution to often intractable problems.

Multi-fidelity Bayesian optimisation integrates function evaluations across a number of different fidelities. 
By applying a cost-adjusted acquisition function, both the subsequent set of decision variables as well as simulation fidelities are selected, accounting for the information/cost trade-off is accounted for.
Incorporating lower fidelity evaluations reduces the time to generate optimal solutions as fewer high-fidelity simulations have to be performed. 
Multi-fidelity approaches contribute to environmental savings in settings where evaluations necessitate substantial computational resources, such as large CFD simulations, reducing life-cycle emissions of overall experimentation, design, and optimisation procedures.
Multi-fidelity Bayesian optimisation has been applied for the design of reactor and tube simulations \cite{savage2023multifidelity}, battery design \cite{folch2023combining}, hyper-parameter optimisation \cite{BOAH}, and the design of ice-sheet simulations \cite{thodoroffmulti}.

\subsubsection{DARTS}

The DARTS framework for the design and analysis of reactor and tube simulations was proposed by \cite{savage2023multifidelity}. 
The approach takes advantage of multi-fidelity simulations and integrates the optimisation process with OpenFOAM within a single Python framework.
At the core of the framework, simulations consisting of a set of decision variables $\mathbf{x}\in\mathcal{X}$ and fidelities $\mathbf{z}\in\mathcal{Z}$ are chosen based on a cost-adjusted acquisition function to solve the following equation
\begin{align}
    \mathbf{x}^* = \arg\max_{x\in\mathcal{X}} f(\mathbf{x},\mathbf{z}_{\bullet})
\end{align}
where $\mathbf{z}_{\bullet}$ is the element-wise vector of highest fidelity parameters\footnote{In the case that the dimensionality of $\mathbf{z}$ is greater than 2 and the set is non-ordered, this is specified as the element-wise maximum of fidelity values.}.
Equation \ref{cost_adjusted} presents the cost adjusted acquisition function, 
\begin{align}
    \mathbf{x}_{t+1},\mathbf{z}_{t+1} = \mathop{\mathrm{argmax}}_{(\mathbf{x},\mathbf{z})\in\mathcal{X}\times\mathcal{Z}} \; \frac{ \mu_{\hat{f}_t}(\mathbf{x},\mathbf{z}_\bullet) + \beta^{1/2}\sigma_{\hat{f}_t}(\mathbf{x},\mathbf{z}_\bullet)}{ \mu_{\lambda_t}(\mathbf{x},\mathbf{z})\sqrt{1-k((\mathbf{x},\mathbf{z}),(\mathbf{x},\mathbf{z_{\bullet}}))^2}}. \label{cost_adjusted}
\end{align}
The criteria balances the exploration, exploitation, and cost of computational experiments and allows for multiple continuous fidelity parameters to be considered. 
The framework also guarantees an evaluated high-fidelity solution to be returned.
Further information regarding the DARTS framework can be found in \cite{savage2023multifidelity}.

We define a maximum time budget approximately depending on the number of parameters in each parameterisation optimised, ranging between 72 and 168 hours in total. 
The DARTS framework involves two hyper-parameters to be decided, $\beta$ defining the exploration level, and $p_c$ which dictates how conservative the stopping criteria is.  
We select these as $\beta=1.5$ and $p_c=2$ based on values that were deemed to be robust from previous work concerning a similar problem.

\subsubsection{Optimisation Objective}

Previous work has demonstrated the optimisation and analysis of coiled tube reactors through. 
The resulting residence time distribution (RTD) from a simulation is approximated using a tanks-in-series model, with a larger number of theoretical tanks representing stronger plug flow performance. 
In these studies, RTDs are observed to have been unimodal and relatively symmetrical. 
Due to the extensive design space applied within this article, we found that simulations can return non-ideal distributions that contain particularly long tails, or are bimodal.
As more complex fluid flows are induced by the reactor designs, the resulting distributions become less ideal, and the tanks-in-series model provides a poor approximation.

Therefore, in this work we propose a composite objective function that serves to not only maximise plug flow performance, but also encourages symmetric and unimodal RTDs, accounting for designs to return non-ideal concentration distributions. 

Equation \ref{n_cost} demonstrates the quantity, denoted $f$, that is minimised during optimisation.
\begin{align}
        \hat{E}_i(N) &= \frac{N(N\theta_i)^{N-1}}{(N-1)!}e^{-N\theta_i} \quad i = 1,\dots,d\\
        N^* &= \arg\min_N \frac{1}{d}\sum_{i=1}^d \left({E_i} - \hat{E}_i(N) \right)^2\\
        f &= \underbrace{\frac{\alpha}{d} \sum_{i=1}^d \left({E_i} - \hat{E}_i(N^*) \right)^2}_{\substack{\text{Weighted}\\\text{MSE}}} - \underbrace{N^*}_{\substack{\text{Equivalent}\\\text{tanks-in-series}}}
        \label{n_cost}
\end{align}

where $d$ is the number of data points contained within an RTD returned from a simulation, $\hat{E}$ and $E$ are the predicted and returned dimensionless concentration values respectively, $\theta$ represents dimensionless time, and $N$ represents tank-in-series.
Based on initial testing, a value of 100 is assigned to $\alpha$, ensuring that the tank-in-series, and the non-ideality error are weighted approximately equally (approximately between 50-100 each).

All code was evaluated using 64 CPUs, a single RTX6000 GPU to aid with training and evaluating Gaussian processes, and 64Gb of RAM. 
All code used within this article can be found within the associated repository \url{https://github.com/OptiMaL-PSE-Lab/pulsed-reactor-optimisation}.
Code containing only the respective parameterisations and function evaluations for benchmarking purposes can be found within the associated repository \url{https://github.com/OptiMaL-PSE-Lab/pulsed-reactor-optimisation}.

\subsection{Experimental Validation}

The selected designs were exported into the Standard Tessellation Language (.STL) file format from their mesh representation and modified resulting into 3D printable models. 
This was achieved by incorporating a bounding cylinder to encompass each reactor. 
The linear sections at the inlet and outlet were specifically configured with 8 mm and 10 mm outside diameter (OD) tube fittings, a requirement for interfacing with the experimental apparatus. 
Finally, the bounding cylinder was reduced with the internal volume between the coil removed to minimise the resin required for printing.
The designs were printed using a FormLabs Form3+ 3D printer, with a Clear V4 resin, following the manufacturer's default settings.
A post-processing phase was performed, comprising of a washing stage in Isopropyl Alcohol (IPA) for 20 minutes, a drying period extending 24 hours, and a concluding post-cure treatment in a FormCure chamber, maintained at a temperature of 60\textsuperscript{o}C for a 30-minute interval.

The RTD method was the same as reported by \citet{McDonough2019a} and applied by \citet{savage2023multifidelity}.
A 0.1 M KCl aqueous tracer solution was injected at the inlet of each reactor, and the conductivity at the outlet was measured throughout time.
The net flow of deionized water and tracer injection were controlled using three separate OEM syringe pumps (C3000, TriContinent) that were hydraulically linked to the reactor via PTFE tubing routed through a custom Swagelok piece.

The concentrations used for the Villermaux–Dushman reaction were 0.0075M H\textsubscript{2}SO\textsubscript{4}, 0.008M KI, 0.0015M KIO\textsubscript{3}, 0.023M NaOH, and 0.0225M H\textsubscript{3}BO\textsubscript{4} (all in mol/L). 
The reactants were introduced into the reactor using separate syringe pumps, and the absorbance at 353 nm caused by tri-iodate ions was measured at the reactor outlet using a UV-Vis spectrometer in a 10 mm pathlength cuvette.
Three repeats were performed for each reactor design to ensure reproducibility.
Figure \ref{polar_gp}c provides an overview of the methodology of this article.

\section{Data Availability Statement}

The OpenFOAM case files for all four reactors presented in \textbf{Figure 1} are made available within the Supplementary Information files, as well as STL files for the reactors presented in \textbf{Figure 3}. The raw experimental data presented in \textbf{Figure 4} is included within the Supplementary Information files. 

\section{Code Availability}
Code used for the closed-loop simulation and optimisation can be located at \url{https://github.com/OptiMaL-PSE-Lab/pulsed-reactor-optimisation}. For use as an optimisation benchmark problem, the reactor simulations are additionally available in the form of a Docker-based REST API with code and instructions at \url{https://github.com/trsav/reactor_benchmark}.

\Urlmuskip=0mu plus 1mu\relax
\bibliographystyle{unsrtnat}
\bibliography{refs} 

\newpage

\end{document}